\documentclass[12pt]{article} 
\usepackage{graphicx}
\usepackage{dcolumn}
\usepackage{amsmath}
\usepackage{array}
\usepackage{bm}
\usepackage{textcomp}

\title{Light and heavy baryon masses: \\ the $1/N_c$ expansion and the quark model }
\author{\textbf{Fabien Buisseret}\thanks{F.R.S.-FNRS Postdoctoral Researcher; fabien.buisseret@umh.ac.be}, 
\textbf{Claude Semay,}\thanks{F.R.S.-FNRS Senior Research Associate; claude.semay@umh.ac.be}\\
Groupe de Physique Nucl\'{e}aire Th\'{e}orique, 
Universit\'{e} de Mons-Hainaut,\\ 
Place du Parc 20,  
B-7000 Mons, Belgium, \\ 
\textbf{Florica Stancu}\thanks{fstancu@ulg.ac.be}, \\
University of Li\`ege,  
Institute of Physics B5, Sart Tilman,\\
B-4000 Li\`ege 1, Belgium, \\
\textbf{Nicolas Matagne,}\thanks{Nicolas.Matagne@theo.physik.uni-giessen.de}\\
Institut f\"{u}r Theoretische Physik, Universit\"{a}t Giessen, \\
D-35392 Giessen, Germany.}

\date{\today}

\begin{document}
\maketitle

\begin{abstract}
We establish a connection between the quark model and
the $1/N_c$ expansion mass formulas used in the description of baryon
resonances. We show that a remarkable compatibility exists between the
two methods in the light and heavy baryon sectors. In particular, 
the band number used to classify baryons in the $1/N_c$ expansion is 
explained by the quark model and the mass formulas for both approaches 
are consistent.  
\end{abstract}


\section{Introduction}

Since pioneering work~\cite{Isgur} in the field, the standard approach for baryon spectroscopy is the
constituent quark model. The Hamiltonian typically contains a spin independent
part formed of the kinetic plus the confinement energies and a
spin dependent part given by a hyperfine interaction. The quark model results are \textit{de facto} model
dependent; it is therefore very important to develop model independent methods that
can help in alternatively understanding baryon spectroscopy and support 
(or not) quark model assumptions. Apart from promising lattice QCD 
calculations~\cite{lattice}, large $N_c$ QCD, or alternatively the $1/N_c$
expansion, offers such a method. 
In 1974 't Hooft generalized QCD from SU(3) to an arbitrary number of colors SU($N_c$) \cite{HOOFT} and suggested a perturbative
expansion
in $1/N_c$, applicable to all QCD regimes. Witten has then applied
the approach to baryons \cite{WITTEN} and this
has led to a systematic and predictive $1/N_c$ expansion method to study static
properties
of baryons. The method is based on the discovery that, in the limit
$N_c \rightarrow \infty$, QCD possesses an exact contracted
SU(2$N_f$) symmetry \cite{Gervais:1983wq} where $N_f$ is the number
of flavors. This symmetry is approximate for finite $N_c$ so that corrections have to be added
in powers of $1/N_c$. Notice that a baryon is a bound state of $N_c$ quarks in the large $N_c$ formalism.

The $1/N_c$ expansion has successfully been
applied
to ground state baryons, either light \cite{DJM94,Jenkins:1998wy} or heavy~\cite{Jenkins:1996de,Jenkins:2007dm}.
Its applicability to excited states is a subject of current
investigations. The classification scheme used in the $1/N_c$ expansion for excited states is based on the standard SU(6) classification as in a constituent quark model. Baryons are grouped into excitation bands $N= 0$, 1, 2,\dots, 
each band containing at least one SU(6) multiplet, the band number $N$ 
being the total number of excitation quanta  in a harmonic oscillator picture.  

The purpose of the present paper is to show that there is
a compatibility between the quark model and the $1/N_c$ expansion methods. 
It is organized as follows. We first give a summary of the $1/N_c$ expansion 
method in Sec.~\ref{barlnc}. Then we present a relativistic quark model in 
Sec.~\ref{barqm} and derive analytic mass formulas from its Hamiltonian 
in Sec.~\ref{massfor}. The comparison between the quark model and the 
$1/N_c$ mass formulas is discussed in Sec.~\ref{compar} and conclusions 
are drawn in Sec.~\ref{conclu}. We point out that the results summarized 
hereafter have been previously presented in Refs.~\cite{us1,us2} for the 
light baryons and \cite{us3} for the heavy baryons. This work aims at  
being a pedagogical overview of these last three references. 

\section{Baryons in large $N_c$ QCD}\label{barlnc}
\subsection{Light nonstrange quarks}
We begin with a summary of the $1/N_c$ expansion in the case $N_f = 2$, 
but the arguments are similar for any $N_f$. The contracted SU(2$N_f$) symmetry 
is here the group SU(4) which has 15 generators: The spin and isospin subgroup
generators $S_i$ and $T_a$ and operators acting on both spin and isospin 
degrees of freedom denoted by $G_{ia}$ ($i,a = 1,2,3$). 

The SU(4) algebra is 
\begin{equation*}
[S_i,T_a] = 0, \quad [S_i,G_{ja}]  =  i \varepsilon_{ijk} G_{ka},\quad [T_a,G_{ib}]  =  i \varepsilon_{abc} G_{ic},	
\end{equation*}
\begin{equation}\label{ALGEBRASU4}
[S_i,S_j]  =  i \varepsilon_{ijk} S_k,\quad [T_a,T_b]  =  i \varepsilon_{abc} T_c,\quad [G_{ia},G_{jb}] = \frac{i}{4} \delta_{ij} \varepsilon_{abc} T_c
+\frac{i}{4} \delta_{ab}\varepsilon_{ijk}S_k.
\end{equation}
In the limit $N_c \rightarrow \infty$ one has  $[G_{ia},G_{jb}]
\rightarrow 0$
which implies the existence of a contracted algebra. These SU(4) generators
form the building blocks of the mass operator, at least in the ground state 
band ($N=0$). For orbitally excited states the generators $\ell^i$ of SO(3), 
as well as the tensor operator $\ell^{(2)ij}$ also appear since the 
symmetry under consideration is extended to SU(4) $\otimes$ SO(3). 

In the $1/N_c$ expansion the mass operator $M$  has general form 
\begin{equation}
\label{massoperator}
M = \sum_{i} c_i O_i,
\end{equation}
where the coefficients $c_i$ 
encode the QCD dynamics and have to be determined from a fit to the existing 
data, and where the operators $O_i$ are SU(4) $\otimes$ SO(3) scalars of the form
\begin{equation}\label{OLFS}
O_i = \frac{1}{N^{n-1}_c} O^{(k)}_{\ell} \cdot O^{(k)}_{SF}.
\end{equation}
Here $O^{(k)}_{\ell}$ is a $k$-rank tensor in SO(3) and $O^{(k)}_{SF}$
a $k$-rank tensor in SU(2)-spin,
but invariant in SU(2)-flavor. 
The lower index $i$ in the left hand side represents a specific
combination.
Each $n$-body operator is multiplied by an explicit factor of
$1/N^{n-1}_c$ resulting from the power counting rules \cite{WITTEN}, where $n$
represents the minimum of gluon exchanges to generate the operator.
For the ground state, one has $k$ = 0. For excited states the $k = 2$
tensor is important. In practical applications, it is customary to include terms
up to $1/N_c$ and drop higher order corrections of order $1/N_c^2$.

As an example, we show the operators
used in the calculation of the masses of the $[{\bf 70},1^-]$ multiplet
up to
order $1/N_c$ included \cite{Matagne:2006dj} (the sum over repeated indices is implicit)
\begin{eqnarray}\label{example}
O_1 = N_c \, {\bm 1} \negthinspace \negthinspace\negthinspace
\negthinspace\negthinspace&, &
~O_2  =  \frac{1}{N_c}\ell^i S^i,
~~~O_3 = \frac{1}{N_c} T^aT^a,
~~~O_4 = \frac{1}{N_c} S^iS^i,\nonumber\\
& O_5 & =  \frac{15}{N_c^2}\ell^{(2)ij}G^{ia}G^{ja},
~~~O_6 =  \frac{3}{N_c^2}\ell^iT^aG^{ia}.
\end{eqnarray}
Note that although $O_5$ and $O_6$ carry a factor of $1/N_c^2$ their matrix elements are
of order $1/N_c$ because they contain the coherent operator $G^{ia}$
which brings an extra factor $N_c$. $O_1 = N_c \, {\bm 1} $ is the trivial operator, proportional
to
$N_c$ and the only one surviving when $N_c \rightarrow \infty$
\cite{WITTEN}. The operators $O_2$ (spin-orbit), $O_5$ and $O_6$ are relevant for orbitally excited
states only. All the SU(4) quadratic invariants
$S^iS^i$, $T^aT^a$ and $G^{ia} G^{ia}$ should enter the mass
formula but they are
related to each other by the operator identity~\cite{Jenkins:1998wy}
\begin{equation}
\label{CASIMIR}
 \left\{S^i,S^i\right\} + \left\{T^a,T^a\right\} + 4  \left\{G^{ia},
 G^{ia}\right\} = \frac{1}{2} N_c (3N_c + 4),
\end{equation}
so one can express $G^{ia} G^{ia}$ in terms of $S^iS^i$ and $T^aT^a$.

Assuming an exact SU(2)-flavor symmetry, the mass formula for the ground state band up to order $1/N_c$ takes the following simple form~\cite{Jenkins:1998wy}
\begin{equation}\label{GS}
 M = c_1 N_c + c_4\frac{1}{N_c}S^2 + \mathcal{O}\left(\frac{1}{N_c^3}
 \right),
\end{equation}
which means that
for $N = 0$ only the operators $O_1$ and $O_4$ (spin-spin) contribute to the
mass. 

Among the excited states, those belonging to the $N = 1$ band,
or equivalently to the $[{\bf 70},1^-]$ multiplet,
have been most extensively studied,
either for $N_f = 2$ (see \textit{e.g.} Refs.~\cite{CGKM,Goity,PY1,CCGL,cohen1})
or for $N_f = 3$ \cite{SGS}. The $N = 2$ band contains the $[{\bf 56'},0^+]$, $[{\bf 56},2^+]$,
$[{\bf 70},\ell^+]$ ($\ell$ = 0, 2), and $[{\bf 20}, 1^+]$ multiplets.
There are no physical resonances associated to $[{\bf 20}, 1^+]$.
The few studies related to the $N = 2$ band concern the
$[{\bf 56'},0^+]$ for $N_f$ = 2 \cite{CC00}, $[{\bf 56},2^+]$
for $N_f = 3$ \cite{GSS}, and $[{\bf 70},\ell^+]$ for
$N_f = 2$ \cite{MS2}, later extended to $N_f = 3$
\cite{Matagne:2006zf}. The method has also been applied \cite{Matagne:2004pm}
to highly excited non-strange and strange
baryons belonging to $[{\bf 56},4^+]$,
the lowest multiplet of the $N = 4$ band \cite{SS94}.

The group theoretical similarity of excited symmetric states
and the ground state makes the analysis of these states simple
\cite{GSS,Matagne:2004pm}. For mixed symmetric states, the situation 
is more complex.
There is a standard procedure which reduces the study of mixed
symmetric states to that of symmetric states.
This is achieved by the decoupling of the baryon into an excited
quark and a symmetric core of $N_c - 1$ quarks.
This procedure has been applied to the $[{\bf 70},1^-]$ multiplet
\cite{CGKM,Goity,PY1,CCGL,cohen1,SGS}
and to the $[{\bf 70},\ell^+]$ ($\ell$ = 0, 2) multiplets
\cite{MS2,Matagne:2006zf}. But it has recently been shown that
the decoupling is not necessary \cite{Matagne:2006dj}, provided
one knows the matrix elements of the SU(2$N_f$) generators
between mixed symmetric states. 
The derivation of these matrix elements is not trivial. For SU(4) they
have been derived by Hecht and Pang \cite{HP} in the context of nuclear 
physics and adapted to quark physics in Ref. \cite{Matagne:2006dj},
where it has been shown that the isospin-isospin term becomes as dominant
in $\Delta$ as the spin-spin term in $N$ resonances.

The derivation of SU(6) matrix elements between mixed symmetric states
$[N_c-1,1]$ is underway \cite{MSnew}.

A detailed description of the problems raised by the standard procedure
\cite{CCGL} of
the separation of a system of mixed spin-flavour symmetry $[N_c-1,1]$ 
into a symmetric core of $N_c-1$ quarks and an excited quark
has been given in Refs. \cite{Matagne:2008fw,Matagne:2008zn}.


\subsection{Inclusion of strangeness}
For light strange baryons ($N_f=3$) the mass operator in the
$1/N_c$ expansion has the general form
\begin{equation}
\label{massoperator2}
M = \sum_{i=1} c_i O_i + \sum_{i=1} d_i B_i,
\end{equation}
where the operators $O_i$ are invariants under SU(6) transformations
and the operators $B_i$ explicitly break SU(3)-flavor symmetry.
In the case
of nonstrange baryons, only the operators $O_i$ contribute, see Eq. (\ref{massoperator}). Therefore
$B_i$ are defined such as their expectation values are zero
for nonstrange baryons. The coefficients
$d_i$ are determined from the experimental data including strange 
baryons.
In Eq.~(\ref{massoperator2}) the sum over $i$ is finite and in practice
it containes the most dominant operators. Examples of $O_i$ and $B_i$ can be 
found in
Refs.~\cite{GSS,Matagne:2006zf,Matagne:2004pm}. 

Assuming that each strange quark brings
the same contribution $\Delta M_s$ to the SU(3)-flavor breaking
terms in the mass formula, we define the total contribution of strange quarks 
as \cite{us2}
\begin{equation}\label{break}
n_s ~\Delta M_s = \sum_{i=1} d_i B_i,
\end{equation}
where $n_s=-{\cal S}$ is the number of strange quarks in a baryon, 
${\cal S}$ being its strangeness.

\subsection{Heavy quarks}

The approximate spin-flavor symmetry for large $N_c$ baryons 
containing light $q = \{u, d, s\}$ and heavy $Q =\{ c, b\}$ quarks
is SU(6)$\times$ SU(2)$_c$ $\times$  SU(2)$_b$,
\emph{i.e.} there is a separate spin symmetry for each heavy flavor.
Over a decade ago the $1/N_c$ expansion has been
generalized to include an expansion   
in $1/m_Q$ and light quark flavor 
symmetry breaking \cite{Jenkins:1996de}. The majority of the currently 
available experimental data concerning heavy baryons is related to ground 
state baryons made of one heavy and two light quarks~\cite{PDG}. 
Such heavy baryons, denoted as $qqQ$ baryons, have been recently reanalyzed 
within the combined $1/N_c$ and $1/m_Q$ expansion \cite{Jenkins:2007dm}, 
and masses in good agreement with experiment have been obtained. A first 
attempt to extend this framework to excited heavy baryons can be found 
in Refs.~\cite{lee} but much work remains to be done in this field.
That is why we focus here on the $N=0$ band for $qqQ$ baryons only.   

Let us first consider that SU(3)-flavor symmetry is exact. In 
this case the mass operator $M^{(1)}$ is a flavor singlet and in the 
combined $1/m_Q$ and $1/N_c$ expansion to order $1/m_Q^2$ 
it takes the following 
form
\begin{equation}\label{mlnc}
M^{(1)}=m_Q {\bm 1}
+\Lambda_{qq}+\lambda_Q+\lambda_{qqQ}.
\end{equation}
The leading order term 
is $m_Q$ at all orders in the $1/N_c$ expansion. Next we have 
\begin{equation}
\Lambda_{qq}=c_0\, N_c\, {\bm 1}+\frac{c_2}{N_c}\, J^2_{qq}, \quad {\rm and} \quad\lambda_Q=N_Q \frac{1}{2 m_Q} \left(c^{'}_0\, 
{\bm 1}+\frac{c^{'}_2}{ N^2_c}J^2_{qq}\right), 
\end{equation}
where $\vec J_{qq}$ is identical to the total spin  $\vec S_{qq}$ 
of the light quark pair when one deals with the $N=0$ band. Note that
$\Lambda_{qq}$ contains the dynamical contribution of the light 
quarks and is independent of $m_Q$ while $\lambda_Q$ gives $1/m_Q$ corrections.
The last term, $\lambda_{qqQ}$, contains the 
heavy-quark spin-symmetry violating operator which reads
\begin{equation}
\lambda_{qqQ}=2 \frac{c^{''}_2}{N_c m_Q}\vec J_{qq}\cdot \vec J_Q,
\end{equation}
where $\vec J_Q$ is identical to the spin $\vec S_Q$ of the heavy quark.

The unknown coefficients $c_0$, $c_2$, $c^{'}_0$, $c^{'}_2$, and  $c^{''}_2$
are functions of $1/N_c$ and of a QCD scale parameter $\Lambda$.
Each coefficient has an expansion in $1/N_c$ where the leading term 
(in dimensionless units) is 
of order unity and does not depend on $1/m_Q$. Thus, 
without loss of generality, by including dimensions,  
one can set $c_0 \equiv \Lambda$.  The quantity $\Lambda$, as well as the other 
coefficients, have to be fitted to the available experimental data. In agreement with Ref.~\cite{Jenkins:1996de}, we take
\begin{align}
\label{largenpar}
c_0 =  \Lambda, \quad c_2 \sim \Lambda,\quad\quad c^{'}_0  \sim   c^{'}_2\sim c^{''}_2  \sim  {\Lambda}^2.   
\end{align}

The inclusion of SU(3)-flavor breaking leads to an 
expansion of the mass operator in the SU(3)-violating parameter $\epsilon$ which contains 
the singlet $M^{(1)}$, an octet $M^{(8)}$, and a 27-plet $M^{(27)}$.
The last term brings contributions proportional to $\epsilon^2$
and we neglect it. For $M^{(8)}$ we retain its dominant contribution 
$T^{8}$ to order $N^0_c$. Then the mass formula becomes
\begin{equation}
\label{breaksim} 
M = M^{(1)} + \epsilon T^{8}.
\end{equation}
The flavor breaking parameter $\epsilon$ is governed by the mass 
difference $m_s - m$ (where $m$ is the average of the $m_u$ and $m_d$ masses) 
and therefore is $\epsilon \sim 0.2$-0.3. It is measured in units of
the chiral symmetry breaking scale parameter $\Lambda_{\chi} \sim 1$ GeV.

\section{Quark model for baryons}\label{barqm}
\subsection{Main Hamiltonian}
The quark model used here to describe baryons aims at capturing the main 
physical features of a three-quark system while keeping the formalism as 
simple as possible in order to get analytical mass formulas. 
It contains: Relativistic kinetic energy for the quarks, $Y$-junction 
confining potential, one-gluon exchange potential and quark 
self-energy contribution added as perturbative terms. Let us now shortly 
describe all these ingredients. 

A baryon, seen as a bound state of three valence quarks, can 
be described, at the dominant order, by the spinless Salpeter Hamiltonian 
$H=\sum^3_{i=1}\sqrt{\vec p^{\, 2}_i+m^2_i}+V_Y$,
where $m_i$ is the bare mass of the quark $i$ and where $V_Y$ is the 
confining interaction potential. We use the bare mass of the quarks in 
the relativistic kinetic energy term as suggested by the field correlator 
method~\cite{simo}, but other approaches, like Coulomb gauge QCD, 
rather favor a running constituent quark mass~\cite{cgqcd}. Although very 
interesting conceptually, the influence of this choice on the mass spectra 
should not be so dramatic than it could have been expected at the first 
glance: First, the bare and constituent heavy quark masses are nearly 
identical. Second, the constituent light quark masses quickly decrease 
at large momentum and become similar to the bare masses; a common limit 
is reached for the excited states. The situation is thus mainly different 
for low-lying $nnn$ baryons ($u$ and $d$ quarks are commonly denoted as $n$), 
where the bare mass $m_n$ can be set equal to 0, but where the constituent 
mass is about 300~MeV~\cite{cgqcd}. However, the strength of additional 
interactions like one-gluon exchange (see next section) can be tuned in 
both cases and lead to final mass spectra which are quite similar. 

Both the flux tube model \cite{CKP} and lattice 
QCD \cite{Koma} support the Y-junction picture for the confining 
potential: A flux tube 
starts from each quark and the three tubes meet at the Torricelli 
(or Steiner or Fermat) point of the 
triangle formed by the three quarks, let us say the $ABC$ triangle. 
This point $T$, located at $\vec x_{T}$, minimizes the sum 
of the flux tube lengths and leads to the following confining potential $V_Y=a \sum^{3}_{i=1} \left|\vec{x}_{i}-\vec{x}_{T}\right|$, where the position of quark $i$ is denoted by $\vec x_i$ and where $a$ is the energy 
density of the flux tubes. If all the angles of $ABC$ are less than 120$^{\rm o}$, then the Toricelli point is such that the angles $\widehat{ATB}$, $\widehat{BTC}$, and  $\widehat{ATC}$ are
all equal to 120$^{\rm o}$. If the angle corresponding to an apex is greater 
than 120$^{\rm o}$, the Toricelli point is precisely at this apex. 

As $\vec x_T$ is a complicated three-body function, 
it is interesting to approximate the confining potential by a more tractable 
form. In the following, we shall use 
\begin{align}
\label{ssh2}
H_R&=\sum^3_{i=1}\sqrt{\vec p^{\, 2}_i+m^2_i}+V_R, \\
\label{pot1}
V_R&=k\,a \sum^{3}_{i=1}\left|\vec{x}_{i}-\vec{R}\right|,
\end{align}
where $\vec{R}$ is the position of the center of mass and  $k$ 
is a corrective factor \cite{Bsb04}. The accuracy of the 
replacement~(\ref{pot1}) has been checked to be very satisfactory 
(better than 5\%) in this last reference provided that the appropriate 
scaling factor is used: $k_0=0.952$ for $qqq$ baryons and $k_1=0.930$ for 
$qqQ$ baryons. For highly excited states, the contribution of the 
configurations in which the Toricelli point is located on one of the quarks 
becomes more and more important, and one could think that the center of mass 
approximation~(\ref{pot1}) is then wrong. But in such cases the angle made 
by the Toricelli point and the other two quarks is larger than $120^{\rm o}$ 
and the center of mass is consequently still close to the true Toricelli point. 
The approximation~(\ref{pot1}), although being less accurate for highly 
excited states, remains however relevant.  

\subsection{Perturbative terms}

Besides the Hamiltonian (\ref{ssh2}),
other contributions are necessary to reproduce the baryon masses. We 
shall add them as perturbations to the dominant Hamiltonian~(\ref{ssh2}). 
The most widespread correction is a Coulomb interaction term of the form
\begin{equation}
\Delta H_{oge}=-\frac{2}{3}\sum_{i<j}\frac{\alpha_{S,ij}}{|
\vec x_i-\vec x_j|},
\end{equation}
arising from one-gluon exchange processes, where $\alpha_{S,ij}$ is the
strong coupling constant between the quarks $i$ and $j$. Actually, one should deal 
with a running form $\alpha_S(r)$, but it would considerably increase the 
difficulty of the computations. Typically, we need two values: 
$\alpha_0=\alpha_{S,qq}$ for a $qq$ pair and $\alpha_1=\alpha_{S,qQ}$ for a 
$qQ$ pair, in the spirit of what has been done in a previous study describing 
mesons in the relativistic flux tube model \cite{tf_Semay}. There it was found that $\alpha_1/\alpha_0 \approx 0.7$ describes rather well the 
experimental data of $q\bar q$ and $Q\bar q$ mesons.   

Another perturbative contribution to the  
 mass is the quark self-energy. This is 
due to the color magnetic moment of a quark propagating through the QCD vacuum. It adds a negative contribution to the hadron masses \cite{qse}. The quark self-energy contribution for a baryon is given by
\begin{equation}
\label{qsedef}
\Delta H_{qse}=-\frac{fa}{2\pi}\sum_i\frac{\eta(m_i/\delta)}{\mu_i},
\end{equation}
where $\mu_i$ is the kinetic energy of the quark $i$, that is $\mu_i=\left\langle \sqrt{\vec p^{\, 2}_i+m^2_i}\right\rangle$,
the average being computed with the wave function of the unperturbed spinless 
Salpeter Hamiltonian~(\ref{ssh2}). The factors $f$ and $\delta$ have been computed in quenched and unquenched 
lattice QCD studies; it seems well established that 
$3 \leq f\leq 4$ and ($1.0 \leq \delta \leq 1.3$)~GeV \cite{qse2}. 
The function $\eta(\epsilon)$ is analytically known; we refer the reader to 
Ref.~\cite{qse} for an explicit formula. It can accurately be fitted by
\begin{alignat}{3}
\label{etaap}
\eta(\epsilon)&\approx 1-\beta \epsilon^2 && \quad\textrm{with} \quad \beta=2.85
&& \quad\textrm{for} \quad 0 \le \epsilon \leq 0.3,\nonumber \\
&\approx \frac{\gamma}{\epsilon^2} && \quad\textrm{with} \quad \gamma=0.79
&& \quad\textrm{for} \quad 1.0 \leq \epsilon \leq 6.0. 
\end{alignat}
Let us note that the corrections depending on the parameter $\gamma$ appear at order $1/m_Q^3$ in the mass formula, so they are not considered in this work.

We finally point out that the quark model we developed in this section is spin 
independent. This neglect of the fermionic nature of the quarks is the reason 
why such a model is often called ``semirelativistic": The implicit covariance 
is preserved, but spin effects are absent. Spin dependent contributions 
(spin-spin, spin-orbit, etc.) typically come from relativistic corrections 
to the one-gluon exchange potential. It is useful to mention that in our 
formalism such potential terms between the quarks $i$ and $j$ should be of 
the form~\cite{simo}
\begin{equation}\label{Vsd}
	V_{ij}\propto(\mu_i\mu_j)^{-1}.
\end{equation}

   
\section{Mass formulas}\label{massfor}

\subsection{The auxiliary field method}
The comparison between the quark model and large $N_c$ mass formulas would be 
more straightforward if we could obtain analytical expressions. To this aim, 
the auxiliary field method is used in order to transform the Hamiltonian~(\ref{ssh2}) into an analytically solvable one \cite{Sem03}. With $\lambda=k\, a$, we obtain
\begin{eqnarray}\label{ham3b}
H(\mu_i,\nu_j)=\sum^3_{j=1}\left[\frac{\vec{p}^{\, 2}_j+m
^2_j}{2\mu_j}+\frac{\mu_j}{2}\right]+\sum^3_{j=1}\left[\frac{ \lambda^2 (\vec{x}_j-\vec{R})^2}{2\nu_j}+
\frac{\nu_j}{2}\right].
\end{eqnarray}
The auxiliary fields, denoted as $\mu_i$ and $\nu_j$, are operators, and 
$H(\mu_i,\nu_j)$ is equivalent to $H$ up to their elimination thanks to the 
constraints
\begin{align}
\label{elim}
\left.\delta_{\mu_i}H(\mu_i,\nu_j)\right|_{\mu_i=\hat\mu_i}&=0\ \Rightarrow\ \hat\mu_{i}=
\sqrt{\vec{p}^{\, 2}_i+m^2_i}, \nonumber \\
\left.\delta_{\nu_j}H(\mu_i,\nu_j)\right|_{\nu_j=\hat\nu_j}&=0\ \Rightarrow\ \hat\nu_{j}=
\lambda|\vec{x}_j-\vec{R}|.
\end{align}
$\left\langle \hat\mu_{i}\right\rangle$ is the quark kinetic energy, and $\left\langle \hat\nu_{i}\right\rangle$ is the energy of one flux tube, the average being computed with the wave function of the unperturbed spinless 
Salpeter Hamiltonian~(\ref{ssh2}). The equivalence relation between Hamiltonians~(\ref{ssh2}) and (\ref{ham3b}) is $H(\hat\mu_i,\hat\nu_j)=H$.

Although the auxiliary fields are operators, the calculations are considerably 
simplified if one considers them as variational parameters. They have then to be eliminated 
by a minimization of the masses, and their extremal values $\mu_{i,0}$ and 
$\nu_{j,0}$ are logically close to 
$\left\langle \hat\mu_{i}\right\rangle$ and 
$\left\langle \hat\nu_{j}\right\rangle$ respectively \cite{Sem03}. This 
technique can give 
approximate results very close to the exact ones~\cite{naro08}. If the 
auxiliary fields are assumed to be real numbers, the Hamiltonian~(\ref{ham3b}) 
reduces formally to a nonrelativistic three-body harmonic oscillator, for 
which analytical solutions can be found. A first step is to replace the quark 
coordinates $\vec x_{i}=\left\{\vec x_{1},\vec x_{2},\vec x_{3}\right\}$ by 
the Jacobi coordinates $\vec x^{\, '}_{k}=\{\vec R,\vec \xi,\vec \eta\,\}$ defined as~\cite{coqm} 
\begin{equation}\label{cmdef}
\vec R=(\mu_{1}\vec x_{1}+\mu_{2}\vec x_{2}+\mu_{3}\vec x_{3})/\mu_{t},\quad {\rm with}\quad \mu_{t}=\mu_{1}+\mu_{2}+\mu_{3},
\end{equation}
and $	\vec \xi \propto \vec x_1-\vec x_2,\  \vec \eta \propto (\mu_1 \vec x_1+\mu_2 \vec x_2)/(\mu_1+\mu_2)-\vec x_3$.

In the case of two 
quarks with mass $m$ and another with mass $m_3$, the mass spectrum of the
Hamiltonian~(\ref{ham3b}) is given by ($\mu_1=\mu_2=\mu$, $\nu_1=\nu_2=\nu$ by symmetry)
\begin{equation}\label{mass1}	
M(\mu,\mu_3,\nu,\nu_3)=\omega_\xi(N_\xi+3  
/2)+\omega_\eta(N_\eta+3/2)+\mu+\nu+\frac{\mu_3+\nu_3}{2}+\frac{m^2}{\mu}+\frac{m^2_3}{2\mu_3},
\end{equation}
\begin{equation}
{\rm where}\quad \omega_\xi=\frac{\lambda}{\sqrt{\mu\nu}},\quad \omega_\eta=\frac{\lambda}{\sqrt{2\mu+\mu_3}}\sqrt{\frac{\mu_3}{\mu\nu}+\frac{2 \mu}{\mu_3\nu_3}}.
\end{equation}
The integers $N_{\xi/\eta}$ are given by $2n_{\xi/\eta}+\ell_{\xi/\eta}$, 
where $n_{\xi/\eta}$ and $\ell_{\xi/\eta}$ are  the radial and 
orbital quantum numbers relative to the variable $\vec \xi/\vec \eta$
respectively. Moreover, $\left\langle \vec \xi^{\, 2}\right\rangle$ and $\left\langle \vec \eta^{\, 2}\right\rangle$ are analytically known. This eventually allows to compute $\left\langle (\vec x_1-\vec x_3)^2\right\rangle$ and $	\left\langle (\vec x_2-\vec x_3)^2\right\rangle$,	which are needed to know the one-gluon exchange contribution.

The four auxiliary fields appearing in the mass 
formula~(\ref{mass1}) have to be eliminated by solving simultaneously the four 
constraints
\begin{alignat}{2}
\label{const}
&\partial_{\mu}M(\mu,\mu_3,\nu,\nu_3)=0, && \quad \partial_{\mu_3}M(\mu,\mu_3,\nu,\nu_3)=0,\nonumber\\
&\partial_{\nu}M(\mu,\mu_3,\nu,\nu_3)=0, && \quad \partial_{\nu_3}M(\mu,\mu_3,\nu,\nu_3)=0.
\end{alignat}
This task cannot be analytically performed in general, but solutions can 
fortunately be found in the case of light and heavy baryons.
\subsection{Light baryons}
Since we do not distinguish between the $u$ and $d$ quarks in our quark model and commonly denote them as $n$, there are only four possible configurations: $nnn$, $sss$, $nss$ and $snn$, that can all be described by the mass formula~(\ref{mass1}). Important simplifications occur by setting $m_n=0$, which is a good approximation of the $u$ and $d$ quark bare masses. However, the non vanishing value for $m_s$ causes Eqs.~(\ref{const}) to have no analytical solution unless a power expansion in $m_s$ is performed. This is justified \textit{a priori} since the strange quark is still a light one. After such a power expansion, the final mass formula reads~\cite{us1}
\begin{align}
\label{M_qqq}
M_{qqq}&=M_0 + n_s\, \Delta M_{0s} \quad\quad (n_s=0,1,2,3), \nonumber \\
M_0 &= 6\mu_0-\frac{2 k_0 a \alpha_0}{\sqrt{3}\mu_0}-\frac{3 f a}{2\pi\mu_0}, \quad
\Delta M_{0s}=\frac{m^2_s}{\mu_0}\left[\frac{1}{2}-\frac{k_  
0 a \alpha_0}{6\sqrt{3} \mu_0^2}
+\frac{f a}{2\pi}\left(\frac{3}{4\mu^2_0}+
\frac{\beta}{\delta^2}\right) \right] , \nonumber \\\mu_0&=\sqrt{\frac{k_0 a(N+3)}{3}}.
\end{align}
The mass formula $M_{qqq}$ depends only on 
$N=N_\xi+N_\eta$. The contribution of terms proportional to 
$N_\xi-N_\eta$, vanishing for $n_s=0$ and 3, was found to be 
very weak in the other cases by a numerical resolution of Eqs.~(\ref{const}).

An important feature of the above mass formula has to be stressed: It only 
depends on $N$ the total number of excitation quanta of the system. But, this 
integer is precisely the band number introduced in large $N_c$ QCD to 
classify the baryon states in a harmonic oscillator picture. Indeed the 
spinless Salpeter Hamiltonian~(\ref{ssh2}) has been transformed into a 
harmonic oscillator by the auxiliary field method and it is thus natural 
that a such band number appears. The great advantage of the auxiliary field 
method is that it allows to obtain analytical mass formulas for a relativistic 
Hamiltonian while making explicitly  the band number used in the large 
$N_c$ classification scheme to appear. The origin of $N$ is thus explained by 
the dynamics of the three-quark system and the comparison with the $1/N_c$ 
mass formulas is therefore possible. 
\subsection{Heavy baryons}

A mass formula for $qqQ$ baryons can also be found from Eq.~(\ref{mass1}). 
An expansion in $m_s$ is still needed to get analytical expressions, but an 
expansion in $1/m_Q$ can also be done since we deal with one heavy quark. 
One obtains~\cite{us3}
\begin{align}
\label{M_qqQ}
M_{qqQ}&=m_Q+M_1 + n_s\, \Delta M_{1s} + \Delta M_{Q}\quad\quad (n_s=0,1,2), \nonumber \\
M_1 &=  4\mu_1 - \frac{2}{3}\left( \alpha_0\sqrt{\frac{k_1 a}{2N_\xi+3}} 
+ 2 \alpha_1\sqrt{\frac{2 k_1 a}{N+3}}\right)-\frac{f a}{\pi\mu_1}, \nonumber \\
\Delta M_{1s}&=\frac{m_s^2}{\mu_1} 
\left[ \frac{1}{2} 
- \frac{1}{12 \mu_1} \left( \alpha_0\sqrt{\frac{k_1 a}{2N_\xi+3}} 
+ 2 \alpha_1\sqrt{\frac{2 k_1 a}{N+3}}\right) 
+ \frac{f a}{2 \pi}\left( \frac{3}{4 \mu_1^2}+\frac{\beta}{\delta^2} \right) \right], \nonumber \\
\Delta M_{Q}&=\frac{k_1 a}{2m_Q} \left[ \left(1-\frac{f a}{2 \pi \mu_1^2}\right)G(N,N_\eta)
-\frac{\alpha_0}{6} \sqrt{\frac{2N_\eta+3}{2N_\xi+3}}\left(\sqrt{\frac{2(2N_\eta+3)}{N+3}}-1\right) 
\right. \nonumber \\
&\phantom{=\frac{k_1 a}{2m_Q} +}\left.+\frac{4 \alpha_1}{3} \frac{2N_\eta+3}{N+3} \right],\nonumber \\
\mu_1&=\sqrt{\frac{k_1 a(N+3)}{2}}, \quad G(N,N_\eta)=\sqrt{2 N_\eta+3}\left( \sqrt{2 (N+3)} - \sqrt{2 N_\eta+3} \right).
\end{align}
At the lowest order in $m_s$ and $1/m_Q$, this mass formula depends only on 
$N$. However, when corrections are added, the mass formula is no longer symmetric 
in $N_\eta$ and $N_\xi$. Is it still possible to find a single quantum number? 
The answer is yes, provided we make the reasonable assumption that an excited 
heavy baryon will mainly ``choose" the configuration that minimizes its mass.  

The dominant correction of order $1/m_Q$ is the term that depends on the function 
$G(N,N_\eta)$. The baryon mass is lowered when $G(N,N_\eta)$ is minimal, 
that is to say for $N_\eta=N$. The analysis of the dominant part of the 
Coulomb term shows that the baryon mass is also lowered in this case. So it is natural to assume that the favored configuration, 
minimizing the baryon energy, is $N_\eta=N$ and $N_\xi=0$. It is also possible to reach the same conclusion by checking that an excitation of type 
$N_\eta$ will keep the baryon smaller in average than the corresponding excitation 
in $N_\xi$. This is favored because of the particular shape of the potential, having for consequence that the more 
the system is small, the more it is light.

As for light baryons, the quark model shows that heavy baryons can be 
labeled by a single band number $N$ in a harmonic oscillator picture. 
A light diquark-heavy quark structure is then favored since the light quark 
pair will tend to remain in its ground state. Note that the diquark picture 
combined with a detailed relativistic quark model of heavy baryons leads to 
mass spectra in very good agreement with the experimental data~\cite{ebert}. 
\subsection{Regge trajectories}

The band number $N$ emerges from the quark model as a good classification 
number for baryons. It is now interesting to focus on the behavior of the 
baryon masses at large values of $N$, \textit{i.e.} for highly excited states. 
In this limit, the formula~(\ref{M_qqq}) gives
\begin{eqnarray}\label{msqr}
M^2_{qqq}&\approx &12\, ak_0(N+3)-\frac{24}{\sqrt 3}ak_0\alpha_0-\frac{16fak_0}{\pi}+6\left[1+
\frac{fak_0\beta}{\pi\delta^2}\right]n_s m_s^2.
\end{eqnarray}
Our quark model thus states that light baryons should follow Regge
trajectories, that is a linear relation $M^2\propto N$, with a common slope, irrespective of the strangeness of the
baryons. The Regge slope of strange and nonstrange baryons is
also predicted to be independent of the strangeness in the $1/N_c$
expansion method~\cite{GM}. Too few experimental data are unfortunately available to check this result. In the heavy baryon sector, the mass formula~(\ref{M_qqQ}) with $N_\xi=0$ and $N_\eta=N$ becomes at the dominant order
\begin{equation}
(M-m_Q)^2=8a\frac{k_1}{k_0}(N+3).
\end{equation}
This model predicts Regge trajectories for heavy baryons, with a slope of $8 a 
k_1/k_0 \approx 7.8  a$ instead of $12 a k_0\approx 11.4\, a$ for light baryons. 

The Regge slope for light baryons is here given by $12ak_0$. However, from 
experiment we know that the Regge slopes for light baryons and light mesons are
approximately equal. For light mesons, the exact value obtained in the
framework of the flux tube model is $2\pi a$, a lower value than the one 
obtained from formula~(\ref{msqr}). This is due
to the auxiliary field method that has been shown to overestimate the 
masses \cite{hyb1}. What can be it done to
remove this problem is to rescale $a$. Let us define $\sigma$ such that
$12ak_0=2\pi \sigma$; then the formula~(\ref{msqr}) is able to reproduce
the light baryon Regge slope for a physical value $\sigma$ of the flux tube 
energy density. The scaling $a=\pi\sigma/(6k_0)$ will consequently be assumed 
in the rest of this paper. 

\section{Large $N_c$ QCD versus Quark Model results}\label{compar}
\subsection{Light nonstrange baryons}
The coefficients $c_i$ appearing in the $1/N_c$ mass operator can be obtained 
from a fit to experimental data. For example, the case $N=0$ is particularly 
simple. Equation~(\ref{GS}) can be applied to $N$ and $\Delta$ baryons. Taking 
$N_c = 3$ together with $M_N = 940$~MeV for $S=1/2$, and
$M_{\Delta} = 1232$~MeV for $S=3/2$, we get
\begin{equation}\label{C2C4}
c_1^{(N=0)} = 289 ~ \mathrm{MeV},~~~~~   c_4^{(N=0)} = 292 ~ \mathrm{MeV}.
\end{equation}
Since the spin-orbit contribution vanishes for $N=0$, no information can be 
obtained for $c_2$. We refer the reader to 
Refs.~\cite{SGS,GSS,MS2,Matagne:2004pm} for the determination of $c_i$ at 
$N > 0$. 
\begin{figure}[ht]
\includegraphics*[width=16cm]{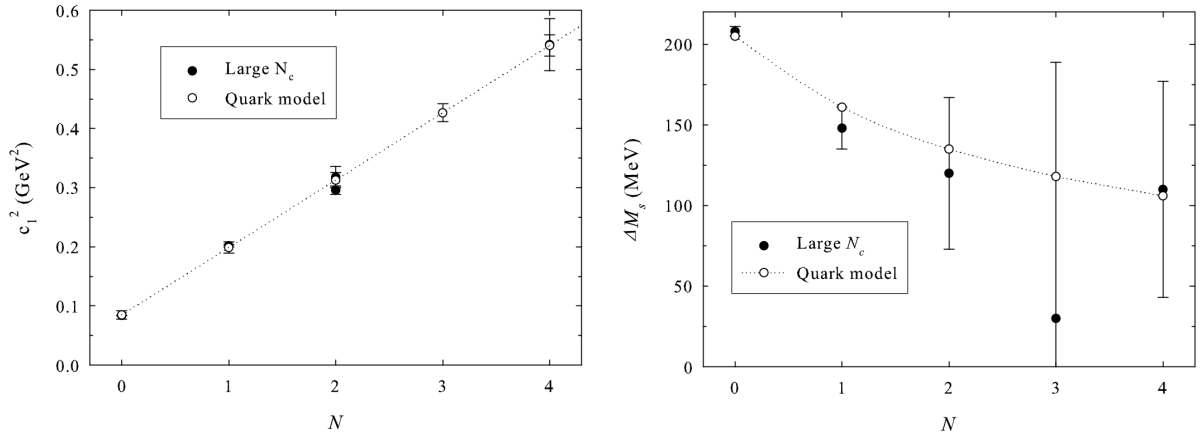}
\caption{Plot of $c^2_1$ (left) and $\Delta M_s$ (right) versus the band number $N$. The values computed in the $1/N_c$ expansion (full
circles) from a fit to
experimental data are compared with the quark model results with $\sigma=0.163$~GeV$^2$, $\alpha_0=0.4$, $f=3.6$, and $m_s=0.240$~GeV (empty circles and dotted line to guide the eyes). No data is available
for $N=3$ in large $N_c$ studies. The large $N_c$ data are nearly indistinguishable from the quark model prediction in the left plot.}
\label{Fig1}
\end{figure}

In the $1/N_c$ expansion method, the dominant term $c_1\, N_c$ in the mass
formula~(\ref{massoperator}) contains the 
spin-independent contribution to the baryon mass, which in a
quark model language represents the confinement and the
kinetic energy. So, it is natural to identify this term with the
mass given by the formula~(\ref{M_qqq}). Then, for $N_c = 3$ we have
\begin{eqnarray}\label{c1qm}
  c^2_1&=&\frac{M^2_{qqq}}{9}=\frac{2\pi}{9}\sigma(N+3)-\frac{4\pi}{9\sqrt 3}\, \sigma\alpha_0-\frac{f\sigma}{3}.
\end{eqnarray}
Figure~\ref{Fig1} shows a comparison between the values of $c^2_1$
obtained in the $1/N_c$ expansion method and those derived from Eq.~(\ref{c1qm}) for various values of $N$. From this comparison
one can see that the results of large $N_c$ QCD are entirely
compatible with the formula~(\ref{c1qm}) provided  $\sigma=0.163$~GeV$^2$, 
a rather low but still acceptable value
according to usual potential models, $\alpha_0=0.4$, and $f=3.6$: These are 
very standard values.

\begin{figure}[ht]
\includegraphics*[width=16cm]{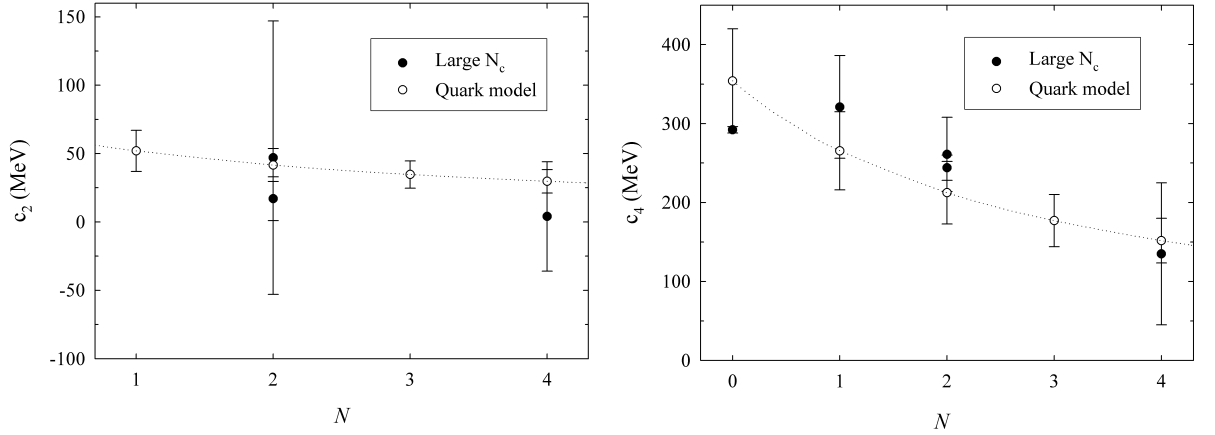}
\caption{
Values of $c_2$ (left) and $c_4$ (right) versus the band number $N$. The values computed in the $1/N_c$ expansion (full circles) from a
fit to
experimental data are compared with results
from formula~(\ref{c2c4qm})
(empty circles and dotted line to guide the eyes). No data is available
for $N=3$ in large $N_c$ studies.}
\label{Fig2}
\end{figure}
Equation~(\ref{Vsd}) implies that
$c_2$ and $c_4\propto\mu^{-2}_0$. Therefore we expect the dependence of $N$
of these coefficients to be of the form
\begin{equation}\label{c2c4qm}
c_2=\frac{c^0_2}{N+3},\quad c_4=\frac{c^0_4}{N+3}.
\end{equation}
We see that such a behavior is consistent with the large $N_c$ results in
Fig.~\ref{Fig2}. We chose $c^0_2=208\pm60$~MeV so that
the point with $N=1$, for which the uncertainty is minimal, is exactly
reproduced. Let us recall that the spin-orbit term is vanishing for $N=0$,
so no large $N_c$ result is available in this case. To compute the
parameter $c^0_4$ a fit was performed on all the
large $N_c$ results. In this way we have obtained $c^0_4=1062\pm198$~MeV.
Note that $c^0_4\gg c^0_2$. This shows that the spin-spin contribution
is much larger than the spin-orbit contribution, which justifies
the neglect of the spin-orbit one in quark model
studies.


\subsection{Light strange baryons} 
We have first to find out the values of $\Delta M_s$ coming from the $1/N_c$ 
expansion. For $N = 0$, 1, and 3, they can be found in Ref.~\cite{GM}, and the 
case $N=4$ is available in Ref.~\cite{Matagne:2004pm}. The situation is 
slightly more complicated in the $N=2$ band due to a larger number of 
available results. We refer the reader to Ref.~\cite{us2} for a detailed 
discussion about the computation of $\Delta M_s$ in this case.

The mass shift due to strange quarks is given in the
quark model formalism by $\Delta M_{0s}$ in Eq.~(\ref{M_qqq}). A comparison of 
this term with its large $N_c$ counterpart
is given in Fig.~\ref{Fig1}, where we used the same parameters as for light 
nonstrange baryons. The only new parameter is the strange quark mass, that we 
set equal to $240$~MeV, a higher mass than the PDG value but rather common in 
quark model studies. One can see that
the quark model predictions are always
located within the error bars of the large $N_c$ results.
Except for $N = 3$, whose large $N_c$ value would actually require further 
investigations, the central values of $\Delta M_s$ in the large
$N_c$ approach are close to the quark model results and they 
decrease slowly and 
monotonously with increasing $N$. Thus, in both approaches, one
predicts a
mass correction term due to SU(3)-flavor breaking which decreases with
the excitation energy (or $N$).


\subsection{Heavy baryons}
As mentioned previously, our present study is restricted to ground state heavy 
baryons made of one heavy and two light quarks. In the $1/N_c$, $1/m_Q$ 
expansion the parameters to be fitted are $\Lambda$, $m_Q$ and $\epsilon\Lambda_\chi$.
At the dominant order, the value of $\Lambda$ can be extracted from the mass combinations~\cite{Jenkins:1996de} 
\begin{equation}\label{mcomb}
\Lambda_Q= m_Q+ N_c \Lambda,\quad
\frac{1}{3} (\Sigma_Q + 2 \Sigma^{*}_Q) - \Lambda_Q
= 2 \frac{\Lambda}{N_c},\quad
\Sigma^{*}_Q - \Sigma_Q =  \frac{3}{2} \left(\frac{2 \Lambda^2}{N_c m_Q}\right),
\end{equation}
resulting from the mass formula (\ref{mlnc}). Here and below the particle label represents its mass. A slightly more complicated 
mass combination, involving light baryons as well as heavy ones, directly 
leads to $m_Q$, that is \cite{Jenkins:2007dm}
\begin{equation}\label{E0}
\frac{1}{3}(\Lambda_Q+2\Xi_Q)-\frac{1}{4}\left[\frac{5}{8}(2N+3\Sigma+\Lambda+2\Xi)-\frac{1}{10}(4\Delta+3\Sigma^*+2\Xi^*+\Omega)\right]= m_Q.
\end{equation}
This mass combination gives 
\begin{subequations}\label{paramLNC}
\begin{equation}\label{mQdef}
	m_c=1315.1\pm0.2~{\rm MeV},\quad m_b=4641.9\pm2.1~{\rm MeV},
\end{equation}
while the value 
\begin{equation}\label{lambdadef}
	\Lambda \approx 324~{\rm MeV}
\end{equation}
\end{subequations}
ensures that the mass combinations~(\ref{mcomb}) are optimally compatible with
the experimental values for $Q = c$ and $b$. A measure of the SU(3)-flavor breaking factor is given by \cite{Jenkins:1996de}
\begin{equation}\label{breakh}
	\Xi_Q-\Lambda_Q=  \frac{\sqrt 3}{2}\, (\epsilon\Lambda_\chi).
\end{equation}
The value $(\epsilon\Lambda_\chi)=206$~MeV leads 
to $\Xi_Q-\Lambda_Q=178$~MeV, which is the average value of the 
corresponding experimental data. 

The new parameters appearing in the quark model are $m_c$, $m_b$, $k_1=0.930$, and $\alpha_1$. For the other parameters we keep the values fitted in the light baryon sector. We take $\alpha_1=0.7\alpha_0$ from the quark model study of 
Ref.~\cite{tf_Semay}. The heavy quark masses can be fitted to the 
experimental data as follows. The quark model mass formula~(\ref{M_qqQ}) is 
spin independent; it should thus be suitable to reproduce the masses of 
heavy baryons for which $J^2_{qq}=0$. Namely, one expects that
\begin{align}
	\left.M_{nnc}\right|_{N=0}=\Lambda_c = 2286.46\pm0.14\ {\rm MeV}, \quad
	\left.M_{nnb}\right|_{N=0}=\Lambda_b = 5620.2\pm1.6\ {\rm MeV}.
\end{align}
These values are reproduced by formula~(\ref{M_qqQ}) with $m_c=1.252$~GeV and 
$m_b=4.612$~GeV. It is worth mentioning that
we predict $\left.M_{nsc}\right|_{N=0}=2433$~MeV and 
$\left.M_{nsb}\right|_{N=0}=5767$~MeV with these parameters.
These values are very close to the experimental $\Xi_c$ and $\Xi_b$ masses respectively.

We can now compare the quark model and the $1/N_c$, $1/m_Q$ mass formulas. 
On the one hand the mass combination~(\ref{E0}) 
leads to $m_c=1315$~MeV and $m_b=4642$~MeV. On the other hand, the quark model 
mass formula~(\ref{M_qqQ}) is compatible with the experimental data provided that $m_c=1252$~MeV and $m_b=4612$~MeV. Both approaches lead to quark masses that 
differ by less than 5\%. Thus they agree at the  
dominant order, where only $m_Q$ is present. 

The other parameter involved in the large $N_c$ mass formula is $\Lambda$. 
A comparison of the spin independent part of the mass formulas (\ref{mlnc}) 
and (\ref{M_qqQ}) leads to the following identification for $N_c = 3$
\begin{eqnarray}  
\label{rel1}
{c_0}= \frac{1}{3}\left.M_1\right|_{N=0}
=
\frac{4}{3}\mu_1 - \frac{2}{27}\sqrt{\frac{k_1 \pi\sigma}{2k_0}} (\alpha_0 
+ 2 \sqrt 2\alpha_1 ) - \frac{f \sigma}{18k_0 \mu_1},\ {\rm with}\ \mu_1=\sqrt{k_1\pi\sigma/4k_0}.
\end{eqnarray}
According to Eqs.~(\ref{largenpar}) and (\ref{lambdadef})
one has $c_0 = \Lambda \simeq 0.324$~GeV.
The quark model gives 0.333~GeV for the 
expression after the second equality sign in Eq.~(\ref{rel1}),
which means a very good agreement for the QCD scale $\Lambda$. The terms of order $1/m_Q$ lead to the identity
\begin{eqnarray} 
\label{rel2}
c^{'}_0 = 2 m_Q \left.\Delta M_Q\right|_{N=0}
=\frac{k_1 \pi\sigma}{6k_0} \left[3\left(\sqrt{2}-1\right) \left(1-\frac{f \sigma}{12k_0  \mu_1^2}\right)
-\frac{\alpha_0}{6} \left(\sqrt{2}-1\right) 
+\frac{4 \alpha_1}{3} \right] . 
\end{eqnarray}
Note that to test this 
relation the value of $m_Q$ is not needed, like for the identity~(\ref{rel1}).
The large $N_c$ parameter, $\Lambda=0.324$~GeV, gives for the left hand side of
(\ref{rel2}) $c^{'}_0\sim \Lambda^2 = 0.096$~GeV$^2$ and the quark model gives 
for the right hand side 0.091 GeV$^2$, which is again a good agreement. Finally, the SU(3)-flavor breaking term is proportional to 
$\epsilon\Lambda_\chi \sim m_s - m\sim m_s$ in the mass formula (\ref{breaksim}). Equations~(\ref{breaksim}), (\ref{M_qqQ}), and (\ref{breakh})
 lead to
\begin{eqnarray}
\frac{\sqrt 3}{2}\epsilon\Lambda_\chi=\left.\Delta M_{1s}\right|_{N=0}
=\frac{m_s^2}{\mu_1} 
\left[ \frac{1}{2} 
- \frac{1}{36 \mu_1}\sqrt{\frac{k_1 \pi\sigma}{2k_0}} \left( \alpha_0 
+ 2 \sqrt 2\alpha_1\right) 
+ \frac{f \sigma}{12k_0}\left( \frac{3}{4 \mu_1^2}+\frac{\beta}{\delta^2} \right) \right].
\end{eqnarray}
The large $N_c$ value $\epsilon\Lambda_\chi=0.206$~GeV and the quark model estimate
$0.170$~GeV also compare satisfactorily. We point out that, except for $m_c$ and $m_b$, all the model parameters 
are determined from theoretical arguments combined with phenomenology, or are 
fitted on light baryon masses. The comparison of our results with the 
$1/N_c$ expansion coefficients $c_0$, $c^{'}_0$ and $\epsilon\Lambda_\chi$ are 
independent of the $m_Q$ values. So we can say that this analysis is 
parameter free. 

An evaluation of the 
coefficients $c_2$, $c_2'$, and $c_2''$ through a computation of the 
spin dependent effects is out of the scope of the present approach. But at the dominant order, one expects from Eq.~(\ref{Vsd}) that $c_2\propto\mu^{-2}_1$ and 
$c_2''\propto\mu^{-1}_1$. The ratio $c_2''/c_2$ should thus be of order 
$\mu_1=356$~MeV, which is roughly in agreement with 
Eq.~(\ref{largenpar}) stating that $c_2''/c_2 \sim \Lambda$. This gives an 
indication that the quark model and the $1/N_c$ expansion method would  
remain compatible if the spin-dependent effects were included.

\section{Conclusions}\label{conclu}

We have established a connection between the quark model and 
the $1/N_c$ expansion both for light baryons and for heavy baryons containing 
a heavy quark. In the latter case the $1/N_c$ expansion is 
supplemented by an $1/m_Q$ expansion due to the heavy quark. A clear correspondence between the various terms appearing 
in the $1/N_c$ and quark model mass formulas is observed, and the fitted coefficients of the $1/N_c$ mass formulas can be quantitatively reproduced by the quark model. 

These results bring reliable 
QCD-based support in favor of the constituent quark model assumptions and lead 
to a better insight into the coefficients $c_i$ encoding the QCD dynamics 
in the $1/N_c$ mass operator. In particular, the dynamical origin of the band 
number labeling the baryons in large $N_c$ QCD is explained by the quark model. 

As an outlook, we mention two important studies that we hope to make in the 
future. First, the $N=1$ baryons of $qqQ$ type are poorly known in the 
$1/N_c$, $1/m_Q$ expansion. They should be reconsidered and compared to the 
quark model. Second, the ground state baryons made of two heavy quarks and a 
light quark could be studied in a combined $1/N_c$, $1/m_Q$ expansion-quark 
model approach, leading to predictions for the mass spectrum of these baryons.  

\section*{Acknowledgement}
F.B. and C.S. thank the F.R.S.-FNRS (Belgium) for financial support.

\end{document}